\documentstyle[pre,aps,multicol,epsf]{revtex}

\begin{document}   

\tightenlines

\title{Coiling and Supercoiling of Vortex Filaments in Oscillatory Media}
\author{Guillaume Rousseau$^{1,2}$, Hugues Chat\'e$^1$, and Raymond Kapral$^3$}
\address{$^1$CEA --- Service de Physique de l'Etat Condens\'e, 
Centre d'Etudes de Saclay, 91191 Gif-sur-Yvette, France\\
$^2$DAEC --- Observatoire de Meudon,
5, place Jules Janssen, 92195 Meudon, France\\
$^3$Chemical Physics Theory Group, Department of
Chemistry, University of Toronto, Toronto, ON M5S 3H6, Canada}

\maketitle

\begin{abstract}
We study the behavior of vortex filaments subject to a uniform density 
of phase twist in oscillatory media described by the 
complex Ginzburg-Landau equation. 
The first instability is a supercritical Hopf bifurcation 
to stable propagating helical vortices.
The secondary instability, 
also a forward Hopf bifurcation, leads to quasiperiodic 
supercoiled filaments. 
The structural changes undergone by these dynamical
objects are akin to those of twisted elastic rods, in spite of the
presence of the  ribbon component of the twist, 
particular to phase singularities.
\end{abstract}


\begin{multicols}{2} 

\narrowtext

Spontaneous oscillations in nonlinear extended media
are commonly observed in nature. Examples include  
reaction-diffusion systems \cite{CWP}, 
certain regimes of fluid flows \cite{MCCPCH}, and 
biological systems with intrinsic clocks \cite{G}. 
The complex Ginzburg-Landau equation (CGLE) \cite{MCCPCH,YK}:
\begin{equation}
\label{cgle}
\partial_t A = A + (1+ i \alpha) \nabla^2 A -  (1+ i \beta) |A|^2 A \;,
\end{equation}
where $A$ is a complex amplitude  describing the slow modulations 
of the oscillations and $\alpha$ and $\beta$ are two real parameters,
plays a central role in 
the analysis of the behavior of oscillatory media, especially
since its validity  extends
far beyond the vicinity of the Hopf bifurcation where it can be 
formally derived. \cite{EXTEND}

The CGLE exhibits the localized structures that often determine  
the spatiotemporal organisation of the medium.
In two dimensions, 
$A=0$ generically at point vortices which
may be sources of spiral waves.
In three space dimensions, the topological defects are
filaments around which ``scroll waves'' can be 
emitted. 
The defects are phase singularities, determined by the topology of 
the phase field $\arg A$. Consequently,
they are not described solely by 
their geometrical structure as given by the Frenet frame at each point 
along the filament; one must also specify how the phase field itself is 
twisted along the vortex (ribbon twist). \cite{TABOR} 

As in lower space dimensions,
the structure and dynamics of three-dimensional (3D) phase singularities is 
best studied in the parameter region where the scroll waves are stable.
Recent simulations and analytical studies in this region \cite{GOG} 
have shown that initially prepared 
untwisted vortex rings shrink and disappear.
Disordered states \cite{AB} and helical structures \cite{ABKp} arising 
from spontaneous stretching and bending of untwisted vortex filaments have 
been found in this region. 

In this Letter, we investigate numerically the effects
of a finite uniform density $\gamma$ of ribbon twist per unit length
on the simplest configuration, a straight infinite filament  
in the regime where both the emitted scroll waves and the 2D
spiral cores are stable. While the limit of small twist was discussed 
in \cite{GOG}, a detailed study of the effects of twist has not 
been carried out for the CGLE, 
even though twisted filaments are easily 
found in the chaotic regimes of the 
3D CGLE \cite{TBP} and have also been observed
in experiments on the Belousov-Zhabotinsky (BZ) 
reaction in the excitable regime. \cite{PAK} 
We find that there is a finite twist density
$\gamma_{\rm c}(\alpha,\beta)$ beyond which the infinite filament
undergoes a supercritical Hopf bifurcation which saturates to produce
helical vortices.
We also characterize the secondary instability, which takes the form
of another Hopf bifurcation leading to the supercoiling of the primary
helices. These two bifurcations are analyzed from a topological viewpoint.
In particular, the writhe is a linear function 
of the excess twist after the 
first instability, and the ribbon twist also varies linearly. 
This suggests analogies between the dynamical structures observed and the 
behavior of twisted elastic rods.

Our results extend similar early observations in excitable media. 
\cite{W-SIAM}
Of special relevance to the present work are studies of helical
solutions of various reaction-diffusion models  
\cite{W-SIAM,HEL-EXCIT,K} which have shown that 
an initially straight vortex filament may adopt a 
helical form with fixed radius for large enough twist. 
Helical filaments have also been observed 
in 3D magnetic resonance images of the BZ reaction. \cite{C}
We believe our findings apply to the excitable case and 
put those results in a more general setting by
elucidating the dynamical nature of the instabilities.
Our work should also lead to an understanding of the behavior of more 
general twisted vortex lines.

The spiral solution of the 2D CGLE is important for the study of 3D
vortex lines: cross sections of scroll waves are spirals, 
up to correction terms calculated in \cite{GOG}. The spiral solution 
is given in polar coordinates $(r,\theta)$ by 
\begin{equation}
\label{spiral}
A_{\rm s}(r,\theta,t) = F(r) \exp i [ \omega_{\rm s} t \pm \theta + \psi(r) ]
\;,
\end{equation}
with $\omega_{\rm s} = -\beta + k_{\rm s}^2 (\beta - \alpha)$, where
$F(r)$ and $\psi(r)$ are two real functions whose asymptotic behavior 
for $r\rightarrow 0$ and $r\rightarrow\infty$ is known, and 
$k_{\rm s}$ is the asymptotic wave number. \cite{H}
In spite of the absence of a general explicit form,
many properties of this solution are known, 
in particular in the parameter region where the asymptotic 
emitted planewave and the discrete core modes are linearly stable. \cite{CORE}
We first consider straight twisted filaments in
this domain of the $(\alpha,\beta)$ plane and use the 
2D stable spiral solution to generate initial conditions efficiently 
by shifting the phase of the spiral regularly
along the third spatial dimension $z$ to create twist: 
$\theta \rightarrow \theta+ \gamma z$.
In the numerical simulations reported below, such initial conditions
have been used in finite boxes of ``height'' $L_z$ with periodic boundary
conditions in $z$, so that the total twist is a multiple of 
$2\pi$: $\gamma L_z = 2\pi n_{\rm t}$. 

The link number Lk, total twist Tw and writhe Wr of the filament
ribbon obey the following law \cite{W,TABOR}:
\begin{equation}
\label{conserv}
{\rm Lk} = {\rm Tw} + {\rm Wr}
={\rm Tw}_{\rm f} +  {\rm Tw}_{\rm r} + {\rm Wr} \quad ,
\end{equation}
where  ${\rm Tw}_{\rm f} = \oint \tau(s) ds$ and
${\rm Tw}_{\rm r} = \oint (d\varphi/ds) \; ds$
are, respectively, the total Frenet and ribbon components of the twist,
defined here in terms of 
local quantities calculated
in the Frenet frame $({\bf t},{\bf n},{\bf b})$ parametrized by the arclength
$s$,  the torsion $\tau$ and the angle $\varphi$
between a given ribbon curve and the binormal vector {\bf b}.
For closed 3D ribbons, the link number Lk is
invariant \cite{TABOR}. In finite boxes periodic in $z$, it is conserved
if the ribbon is not ``cut'', i.e. no reconnection occurs, either of
the filament with itself or with another filament, and we have, simply:
\begin{equation}
{\rm Lk} =2\pi\, n_{\rm t} = \gamma L_z \quad .
\end{equation}
The constraint imposed by the conservation law has important 
consequences since it implies that
filaments with different $\gamma$ values are not continuously 
related; thus, one must consider $\gamma$ as an independent parameter.

\begin{figure}
\centerline{\epsfxsize=7.5truecm
\epsffile{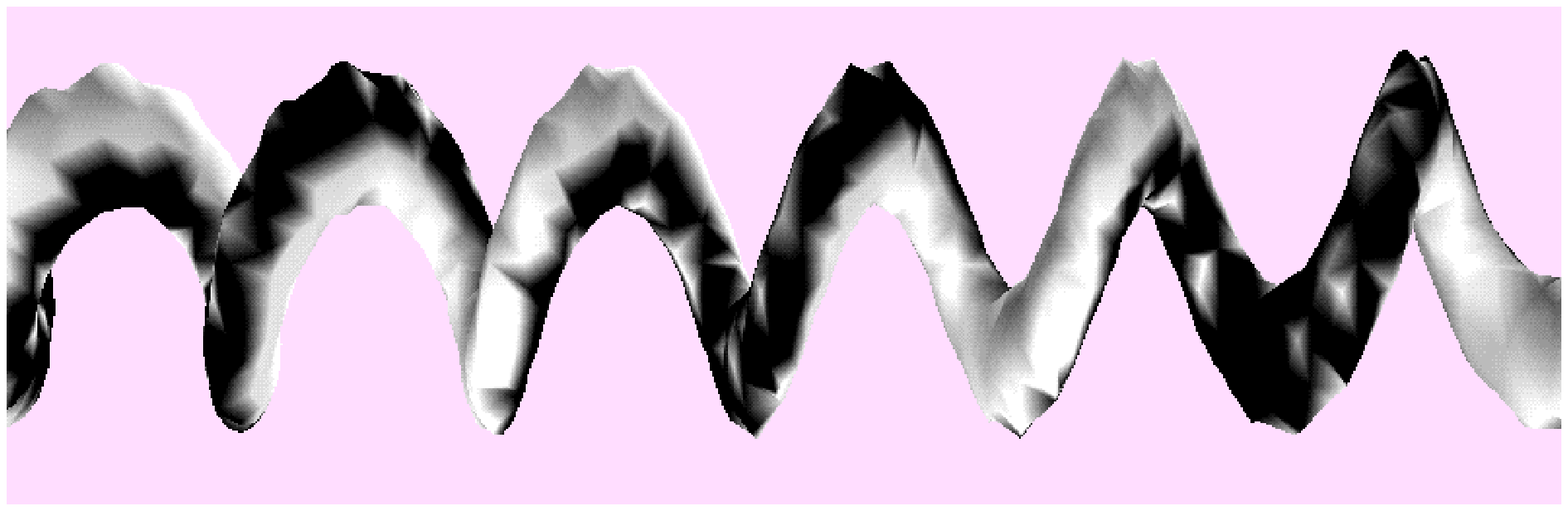}
}
\vspace{-1.5cm}
\centerline{\epsfxsize=5.5truecm
\epsffile{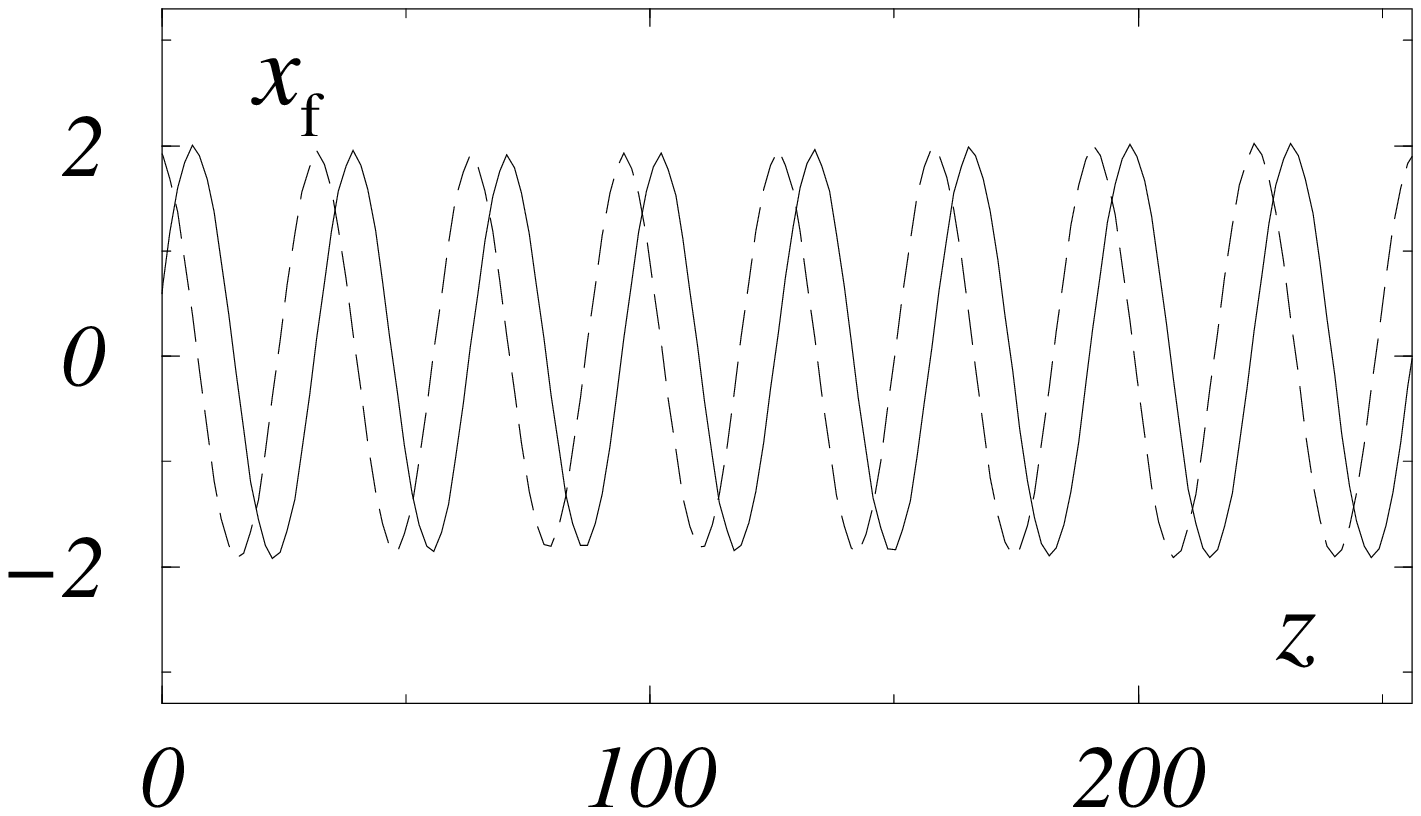}
\hspace{-1.3cm}
\epsfxsize=5.5truecm
\epsffile{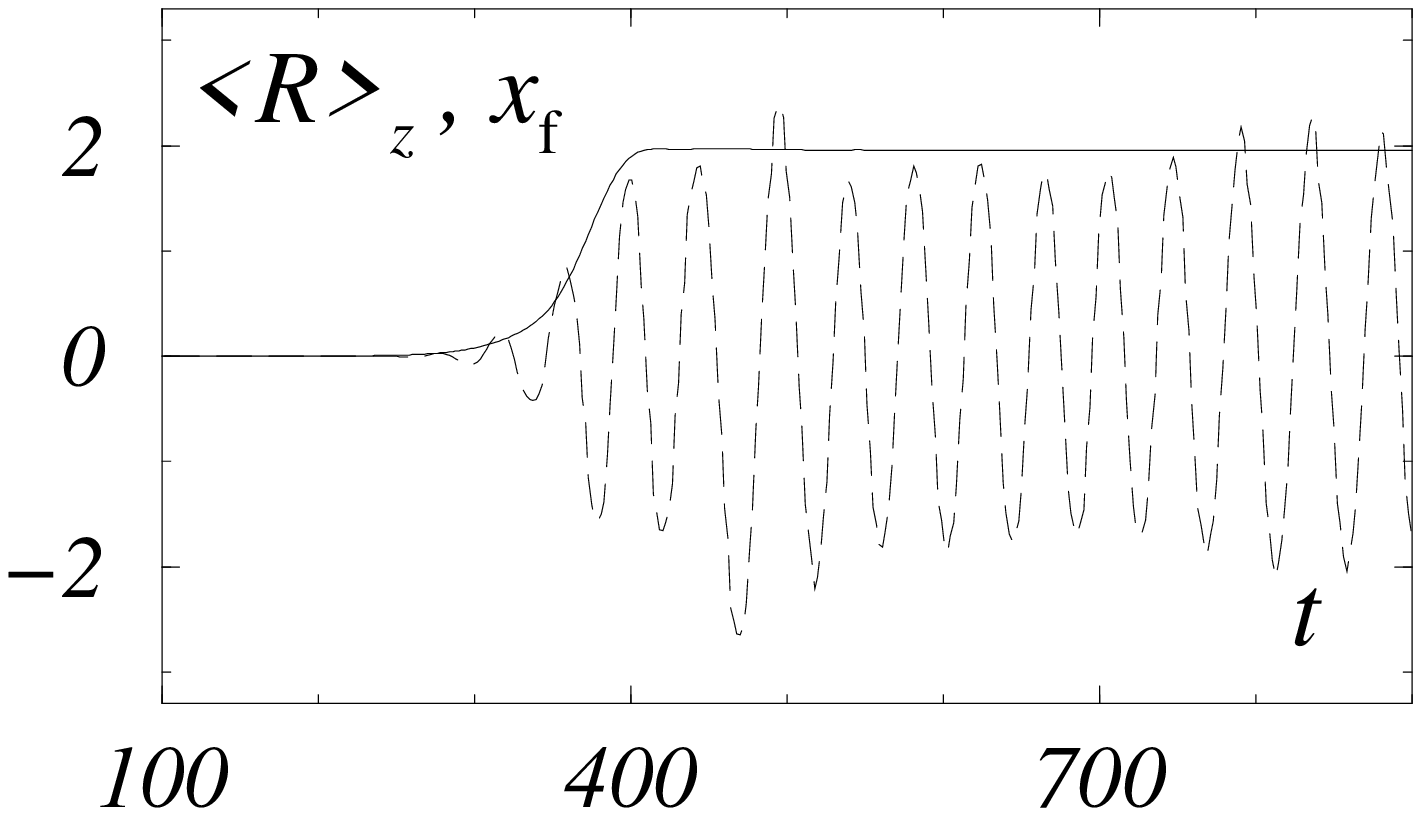}}
\vspace{-0.5cm}
\caption{Stable helical vortex for
 $\alpha=1.7$, $\beta=0$, $\gamma=0.098$ and $L_z=512$ ($L_x=L_y=128$). 
Top: perspective view of the 3D isosurface $|A|=0.6$ 
colored by the phase field ($2\pi$-periodic grey scale). Note the presence of 
ribbon twist ($\gamma_{\rm h}=0.196$, ${\rm Tw}_{\rm r}/L_z=-\gamma$).
Bottom left: $(x,z)$ projection of the filament at $t=2495$ (dashed line)
and $t= 2500$ (solid line).
Bottom right: growth of $\langle R\rangle_z$ (solid line) 
and time series of the $x$
coordinate (dashed line) of a point on the filament at a given~$z$.}
\label{f1}
\end{figure}

Consider a single twisted filament,
generated as described above, evolving under the CGLE. 
Our simulations employed no-flux boundary
conditions in the $x$ and $y$ directions while the box was
periodic in $z$, the axis of the filament. All numerical
results have been checked by changing the box size and 
the spatiotemporal mesh size.

For small $\gamma$,
the solution quickly relaxes to take into account the wavenumber and frequency
shifts \cite{GOG} between the 2D spiral waves and the 3D scroll waves. 
This (stable) solution of the 3D CGLE is $A_{\rm f}(r,\theta,z,t)$:
\begin{equation}
\label{fil}
A_{\rm f} =\sqrt{1-\gamma^2} F(r')  \exp i 
[ \omega_{\rm f} t \pm (\theta+\gamma z) + \psi(r') ] \quad ,
\end{equation}
where $r'=r\sqrt{1-\gamma^2}$ and $\omega_{\rm f}=\omega_{\rm s}(1-\gamma^2)
-\gamma^2\alpha$, with $\omega_{\rm s}$, $F$, and $\psi$ given by the 2D
spiral solution (\ref{spiral}). We have confirmed that our numerical
solutions coincide with the above exact solution when there is no core 
instability.

\begin{figure}[htbp]
\vspace{-0.5cm}
\centerline{\epsfxsize=5.5truecm
\epsffile{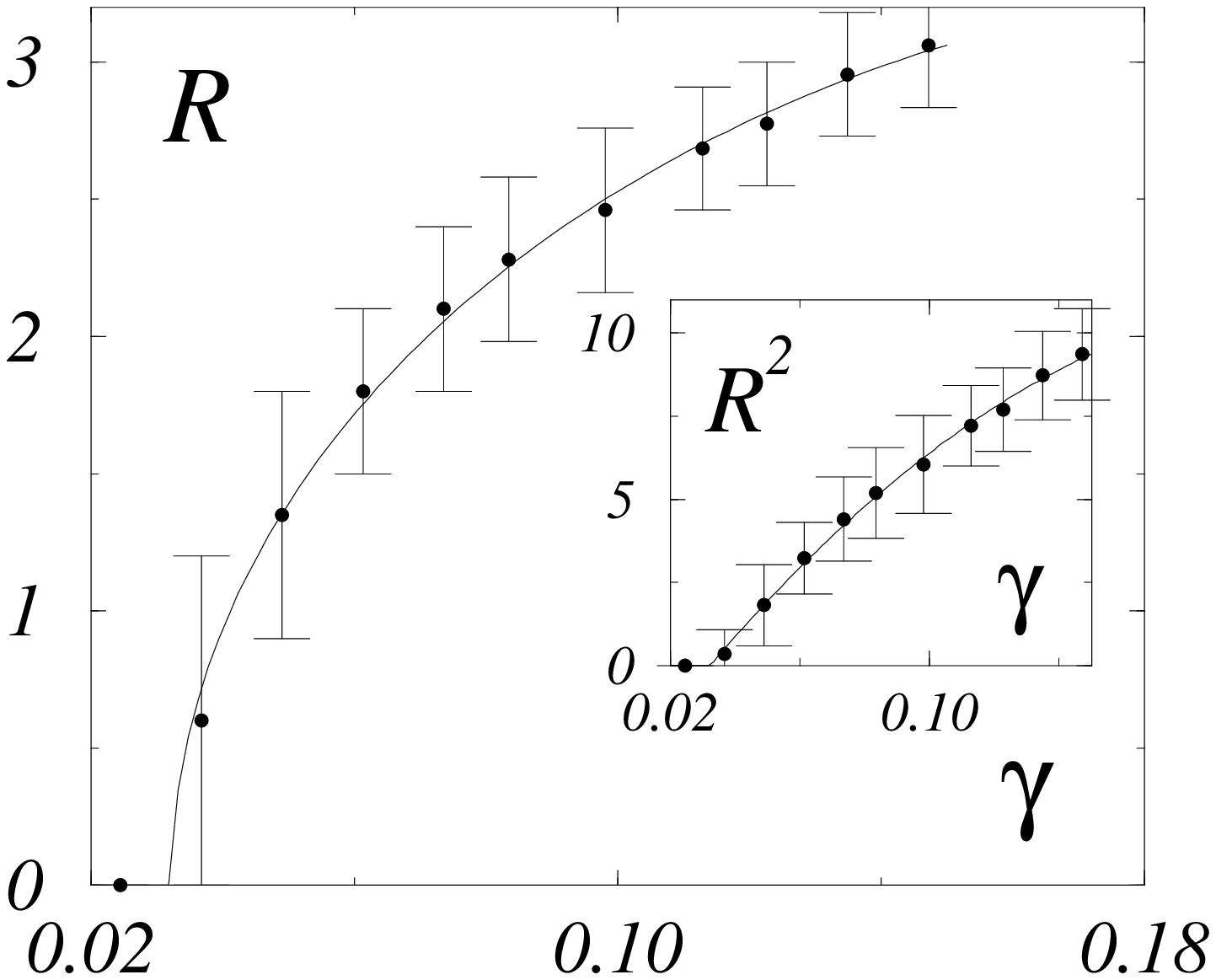}
\hspace{-1.3cm}
\epsfxsize=5.5truecm
\epsffile{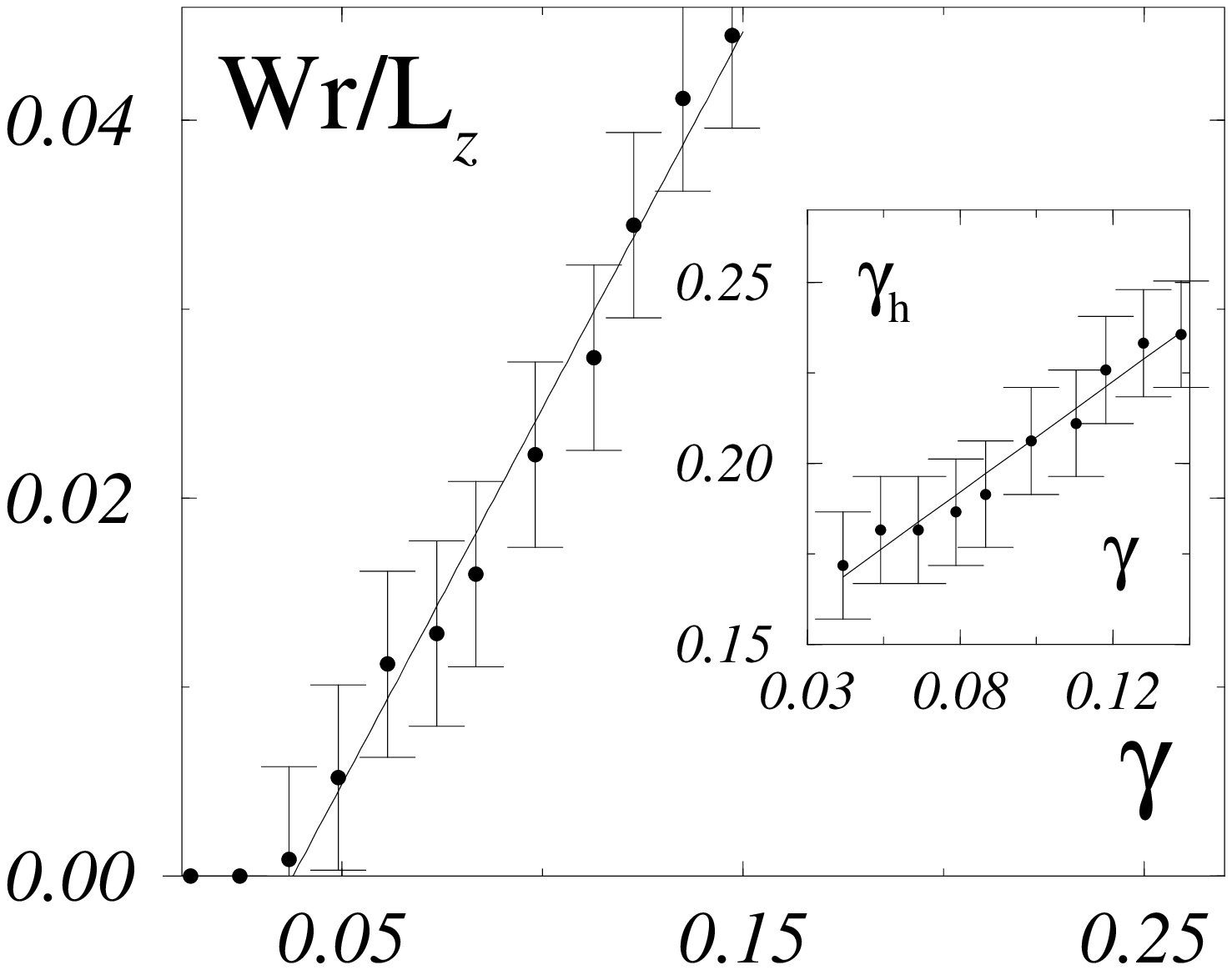}}
\vspace{-0.5cm}
\caption{$\alpha=1.7$ and $\beta=0$: $R(\gamma)$ (left) and 
${\rm Wr}(\gamma)/L_z$ (right)
($\gamma_{\rm c}\simeq 0.032$).
Insets: $R^2(\gamma)$ (left) and $\gamma_{\rm h}(\gamma)$ (right).
}
\label{f2}
\end{figure}

\begin{figure}[htbp]
\vspace{-0.5cm}
\centerline{\epsfxsize=5.5truecm
\epsffile{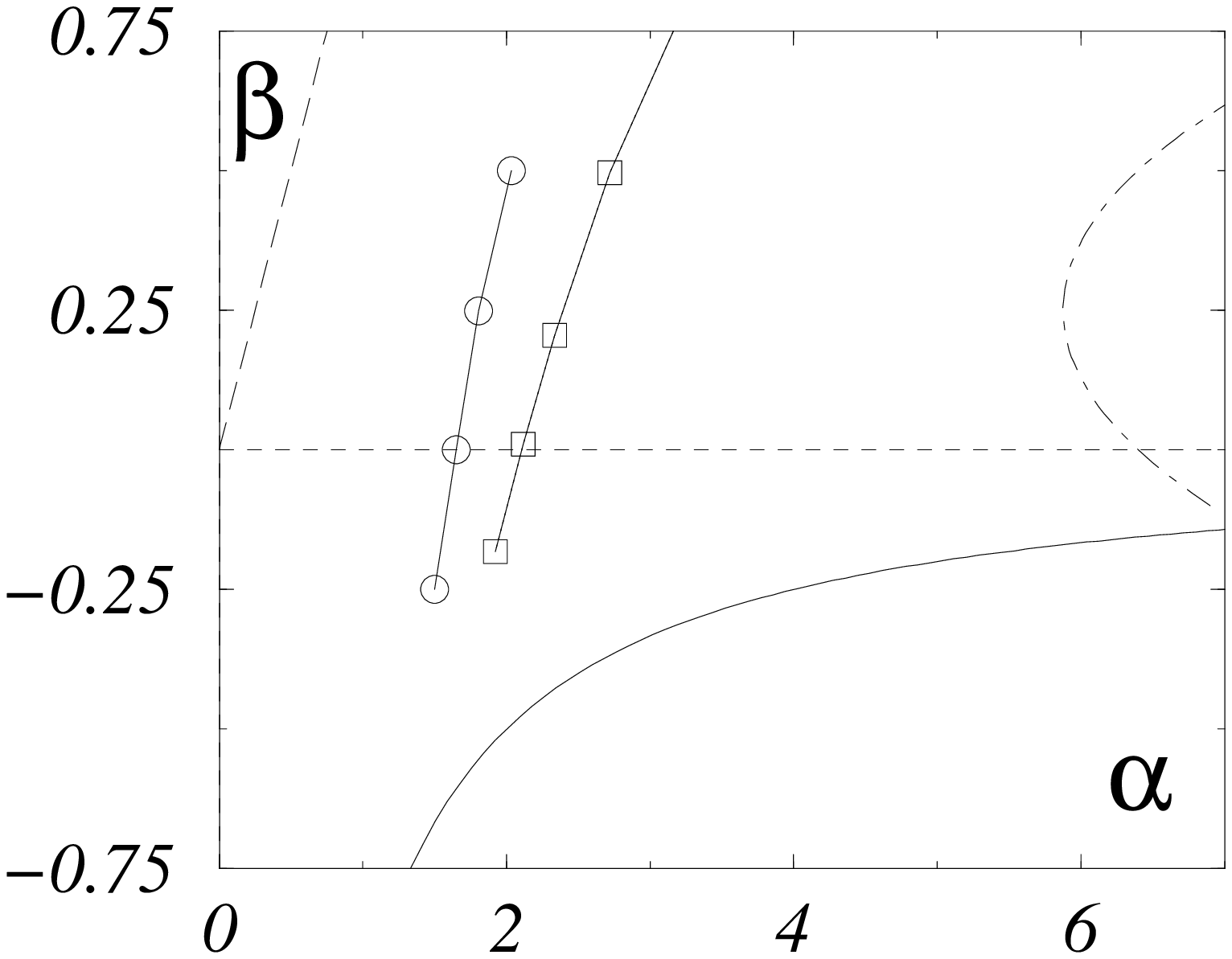}
\hspace{-1.4cm}
\epsfxsize=5.5truecm
\epsffile{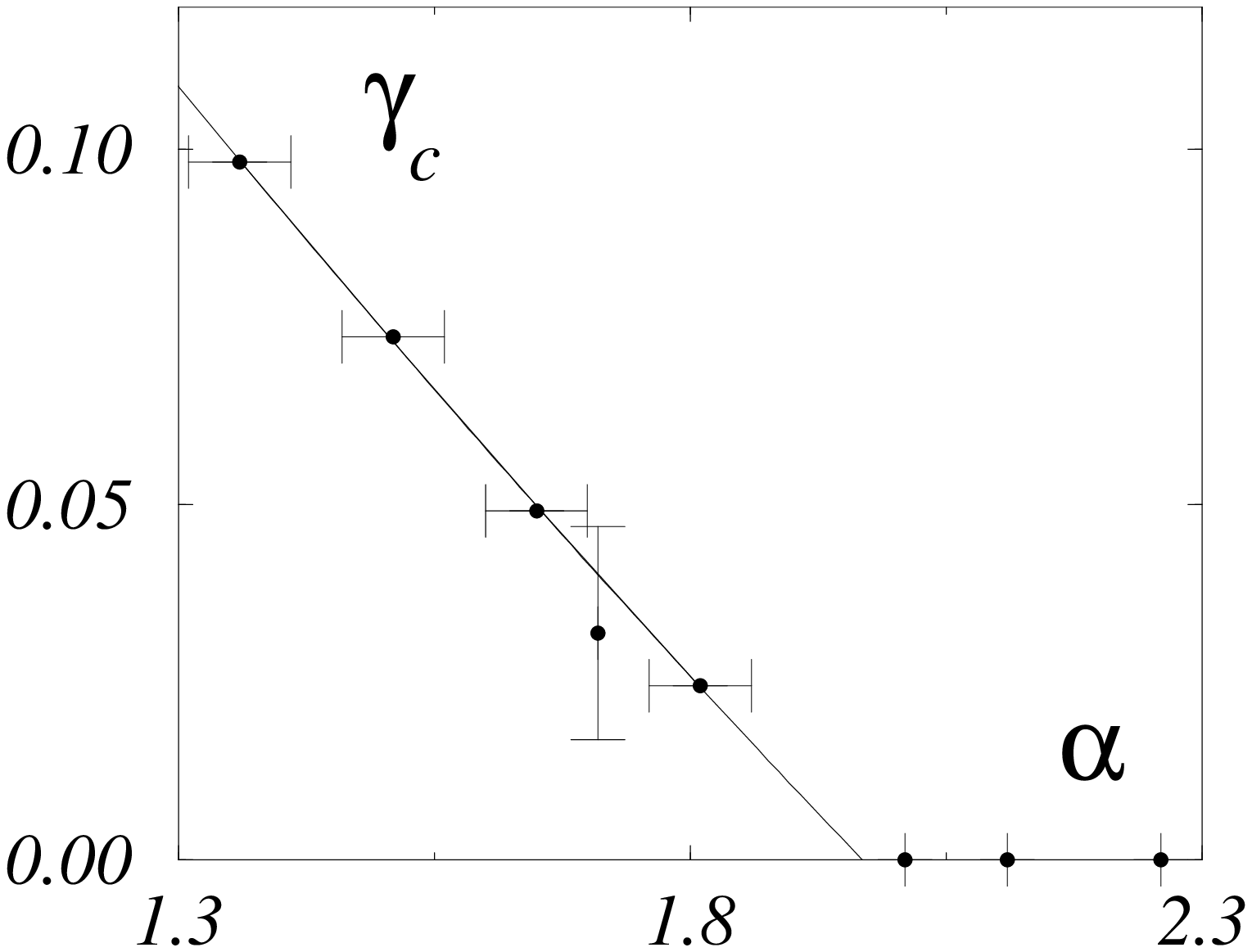}}
\vspace{-0.5cm}
\caption{Left: phase diagram of the 3D CGLE. 
The region of interest lies
between the 2D spiral core instability (dash-dotted) line, the 
phase instability $1\!+\!\alpha\beta\!=\!0$ (solid line), and
the $\alpha=\beta$ line (dashed) where the spiral core is infinite. 
The thin lines are
the $\gamma_{\rm c}= 0.049$ (circles) and $\gamma_{\rm c}=0$ (squares,
from \protect\cite{ABKp})
level curves of the $\gamma_{\rm c}(\alpha,\beta)$ surface.
Right: $\gamma_{\rm c}(\alpha)$ for $\beta=0$. A quadratic fit gives
$\gamma_{\rm c}=0$ for $\alpha=1.95(15)$.} 
\label{f3}
\end{figure}

For large enough $\gamma$ this relaxation is followed by the
growth of periodic modulations along the filament, which becomes a helix
of pitch $\lambda_{\rm h}$ and well-defined radius $R$  (Fig.~\ref{f1}). 
The helix 
rotates with constant angular velocity $\omega$ and any point on the 
helix moves along its axis at a constant velocity 
$v=\omega/\gamma_{\rm h}$ with $\gamma_{\rm h}\equiv 2\pi/\lambda_{\rm h}$.
These results indicate that the straight filament
has undergone a Hopf bifurcation to traveling waves. 
The direction of propagation and rotation is determined by the sign of 
$\gamma$, which breaks the $z\to -z$ parity symmetry.
Varying $\gamma$ continuously across the instability threshold 
$\gamma_{\rm c}$, the behavior
expected for a supercritical bifurcation is observed: 
$R^2 \sim (\gamma-\gamma_{\rm c})$,
while $v$ and $\lambda_{\rm h}$ vary smoothly   (Fig.~\ref{f2}).
The forward nature of the bifurcation is confirmed by
the growth transient (Fig.~\ref{f1} bottom right).
Similarly, keeping $\gamma$ constant and varying $(\alpha,\beta)$ toward
the 2D core-instability line, a bifurcation point is passed and 
the expected small-$R$ behavior 
(e.g. $R^2 \sim (\alpha-\alpha_{\rm c})$ at fixed $\beta$) is observed 
(not shown). 
This suggests the existence of a critical surface 
$\gamma_{\rm c}(\alpha,\beta)$ delimiting the stability of straight
twisted filaments. 
Our  simulations yielded the approximate location 
of the $\gamma_{\rm c}= 0.049$ line in the $(\alpha,\beta)$ plane 
(Fig.~\ref{f3} left). The complete $\gamma_{\rm c}(\alpha,\beta)$
surface could be determined from a  
linear stability analysis of solution (\ref{fil}), 
possibly along the lines of \cite{ABKp},
a rather difficult task left for future work \cite{TBP}.

\begin{figure}[htbp]
\vspace{-0.cm}
\centerline{\epsfxsize=7truecm
\epsffile{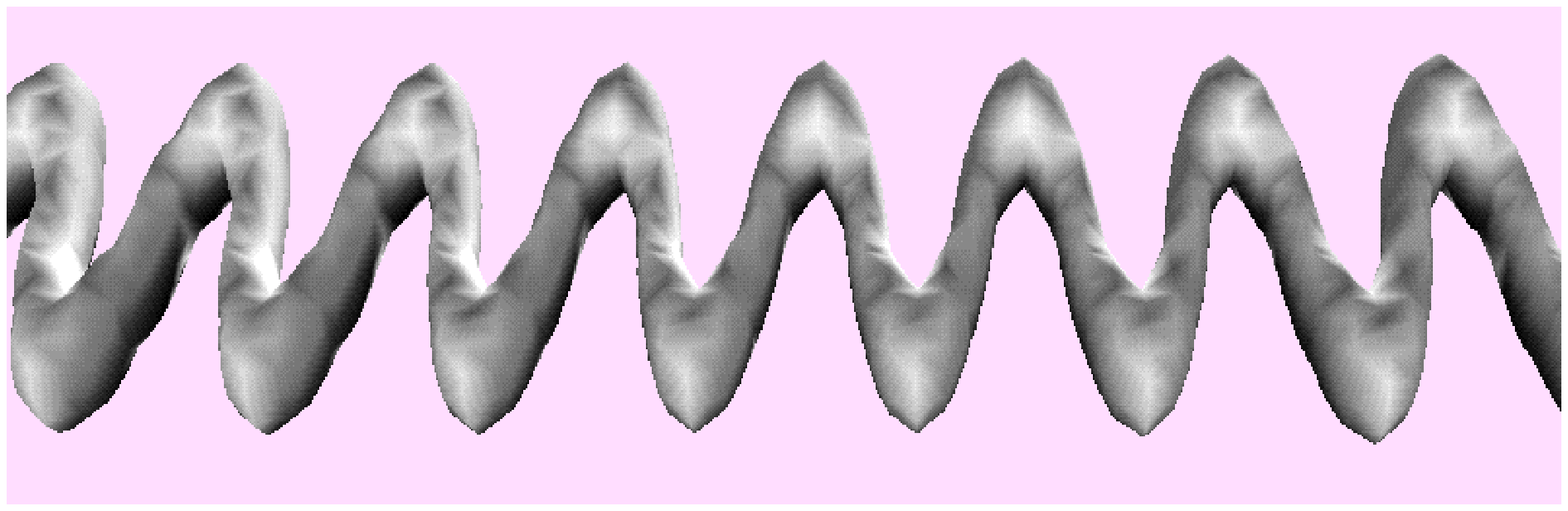}}
\vspace{0.2cm}
\centerline{\epsfxsize=7truecm
\epsffile{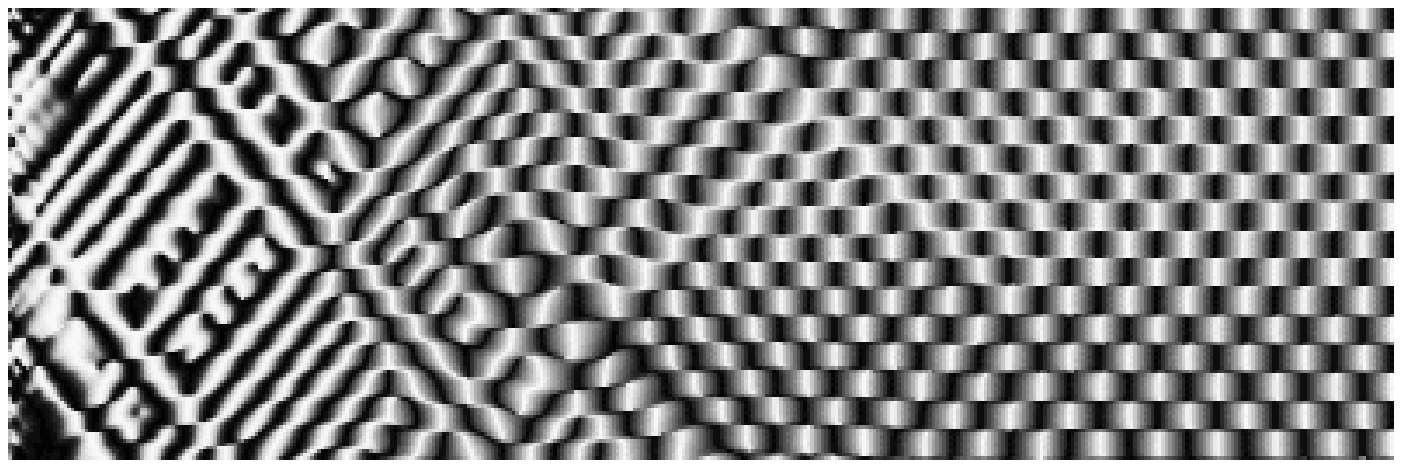}}
\vspace{0.2cm}
\caption{Asymptotic state of an unstable untwisted filament 
($\alpha\!\!=\!\!2.5$, $\beta\!\!=\!\!0$). 
Top: 3D isosurface $|A|\!=\!0.6$ 
colored by the phase field.
Bottom: space-time plot ($t$ running from left to right) of the phase
$\phi(z,t)=\arctan(y_{\rm f}(z)/x_{\rm f}(z))$ where $(x_{\rm f}(z),
y_{\rm f}(z), z)$ parametrize the filament. 
($2\pi$-periodic grey scales $0\!=$black, $\pi\!=$white, were used). 
}
\label{f5}
\end{figure}

The $\gamma_{\rm c}=0$ line determines 
the stability domain of the untwisted filament. In \cite{AB},
it was argued that this solution, which is just the trivial extension
of the 2D spiral, is unstable in the region of the 2D core instability.
It was further argued recently that the instability region of the
untwisted filament extends beyond the 2D core line \cite{ABKp}.
To attack this problem from the viewpoint of twisted filaments, 
we have performed simulations for $\beta=0$
to determine $\gamma_{\rm c}(\alpha)$ (Fig.~\ref{f3}
right). We find, by extrapolation,  $\gamma_{\rm c}=0$ for
$\alpha = 1.95(15)$, in agreement with \cite{ABKp}.
Our results indicate that the instability of the $\gamma=0$ filaments
does {\it not} lead to helices. In this special case, the $z\to -z$
parity symmetry is preserved and one expects, a priori, the superposition
of opposite traveling waves at the linear stage. At the nonlinear level,
we observe the saturation of the two corresponding amplitudes at the same
value and the appearance of a flat, $z$-periodic filament rotating
uniformly around the $z$ axis, the equivalent of standing waves in the 
amplitude representation (Fig.~\ref{f5}).

We now consider our results at non-zero twist from a topological viewpoint.
Since no reconnection occurs across the transition, Lk remains constant.
Given $\lambda_{\rm h}$ and $R$,
all terms of conservation law (\ref{conserv}) can be calculated easily.
For the straight filament, the Frenet frame is degenerate;  
${\rm Tw}_{\rm f}$ and ${\rm Tw}_{\rm r}$ cannot be distinguished.
For the stable helix,
on the other hand, the ribbon twist is given by
${\rm Tw}_{\rm r}/ {\rm Lk} = 1 - \gamma_{\rm h}/\gamma$. 
Moreover,  $\kappa$ and $\tau$ are constant along the $z$ axis: 
$\kappa/(R\gamma_{\rm h}) = \tau = 
\gamma_{\rm h}/(1+R^2\gamma_{\rm h}^2)$.
We then have  \cite{HEL-EXCIT}:
\begin{equation}
\label{modcon}
\frac{{\rm Tw}_{\rm f}}{\rm Lk} = \frac{\gamma_{\rm h}}{\gamma} -
\frac{\rm Wr}{\rm Lk} = 
\frac{\gamma_{\rm h}}{\gamma\sqrt{1+ R^2 \gamma_{\rm h}^2}} \; .
\end{equation}
Exploring a large part of the $(\alpha,\beta)$ domain of interest,  
we consistently found that $\lambda_{\rm h}$ is larger than $\lambda_{\rm s}$, 
the wavelength of the 2D spiral, 
but of the same order, and that the two wavelengths vary similarly with
$(\alpha,\beta)$ \cite{TBP}. Thus the typical scale of the 3D instability 
depends essentially on the wavenumber selected by 2D spiral solution.
We also find that $\gamma_{\rm h} > \gamma$, which implies 
that ${\rm Tw}_{\rm r}<0$. The ribbon
twist ``compensates'' the strong torsion due to the instability. 
For $(\alpha=1.7,\beta=0)$, $\lambda_{\rm h}$ decreases with increasing
$\gamma>\gamma_{\rm c}$ (Fig.~\ref{f2}, right), and thus 
the magnitude of the ${\rm Tw}_{\rm r}$ increases. In fact, our data 
indicates a linear variation with
$\gamma$, but we cannot rule out a weakly quadratic dependence.
Moreover, in agreement with (\ref{modcon}) for small 
$R^2 \sim \gamma - \gamma_c$,
${\rm Wr} \sim (\gamma-\gamma_{\rm c})$, but, remarkably, 
this behavior also extends far from $\gamma_{\rm c}$ (Fig.~\ref{f2}).
(Again, a weakly quadratic dependence is still possible.) 
Thus, all topological quantities seem to vary linearly above threshold,
and, in particular, a {\it constant proportion} of the excess twist 
$\gamma-\gamma_{\rm c}$
is converted into writhe. This is akin to the behavior of elastic rods
\cite{TABOR,ROD}.

\begin{figure}
\centerline{\epsfxsize=8.truecm
\epsffile{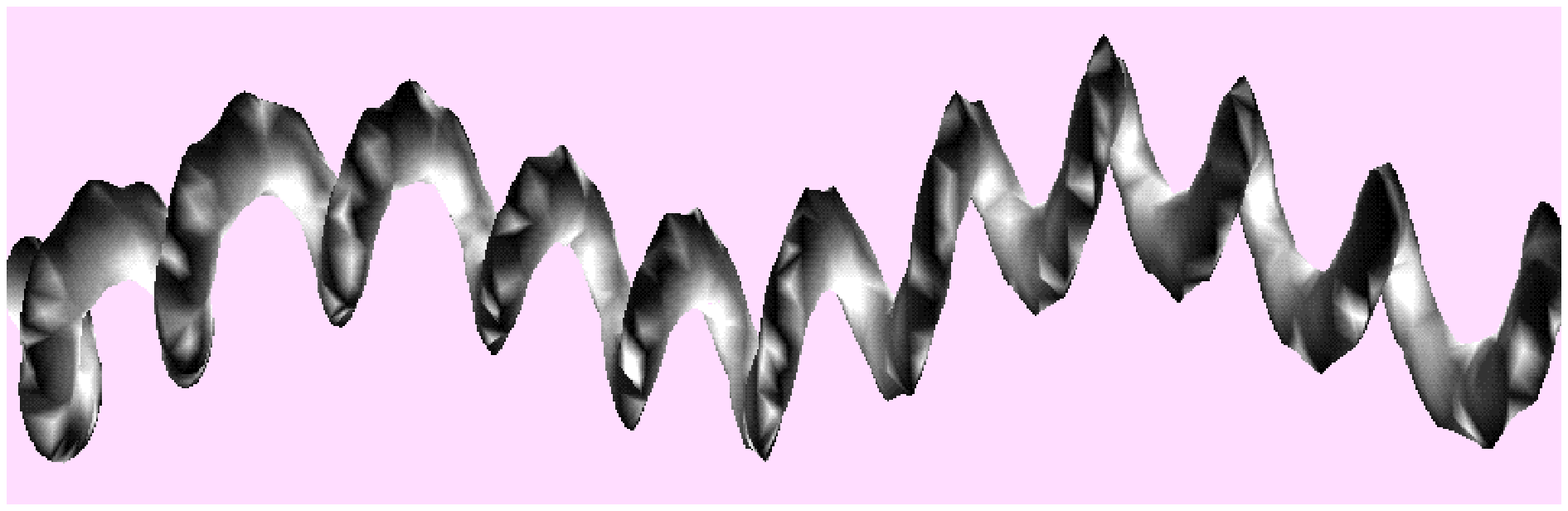}
}

\vspace{-0.4cm}

\hspace{-1.cm}
\epsfxsize=6.2cm
\epsffile{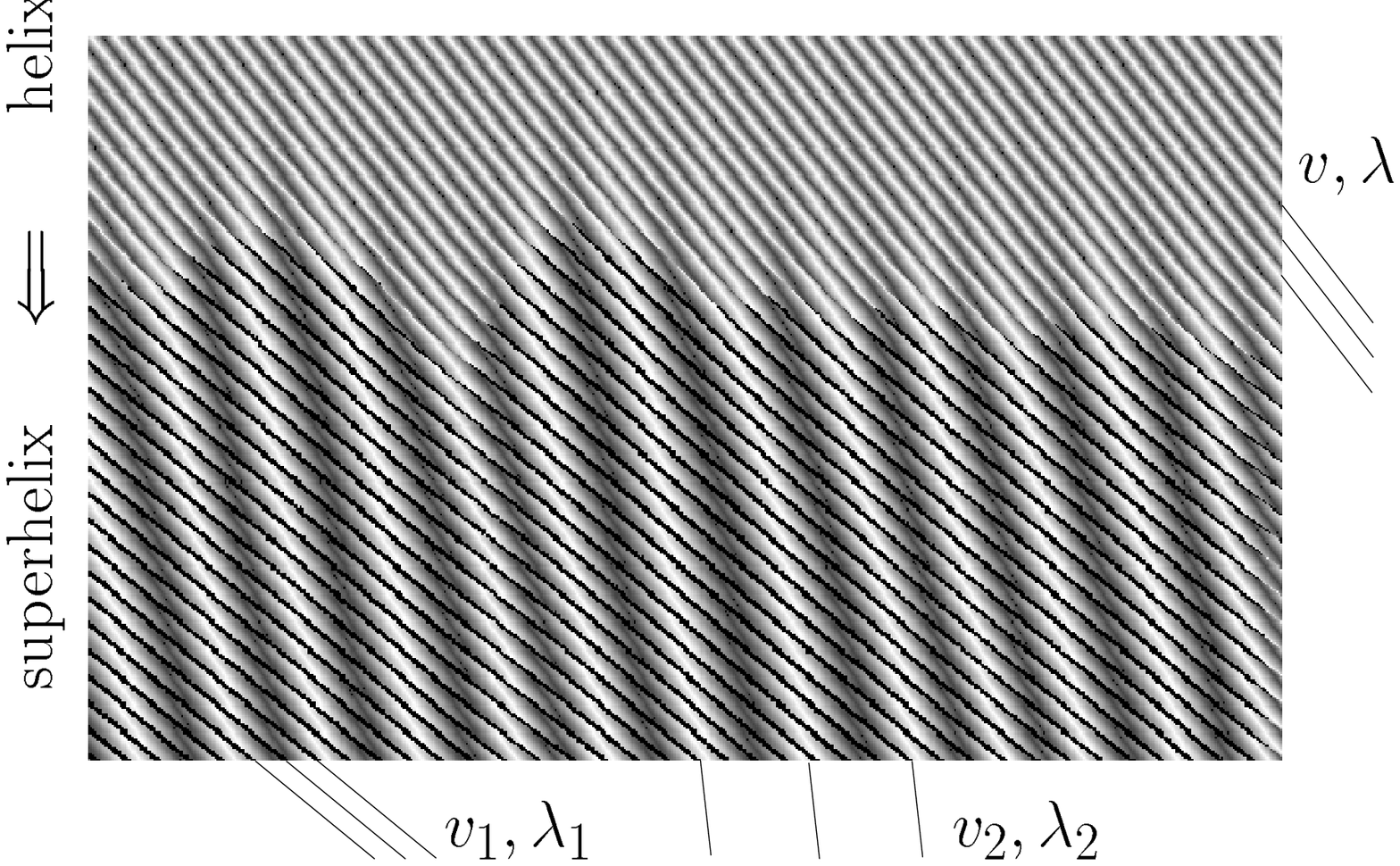} \hfill

\vspace{-8.5cm}

\hspace{4.6cm}
\epsfxsize=4.5truecm
\epsffile{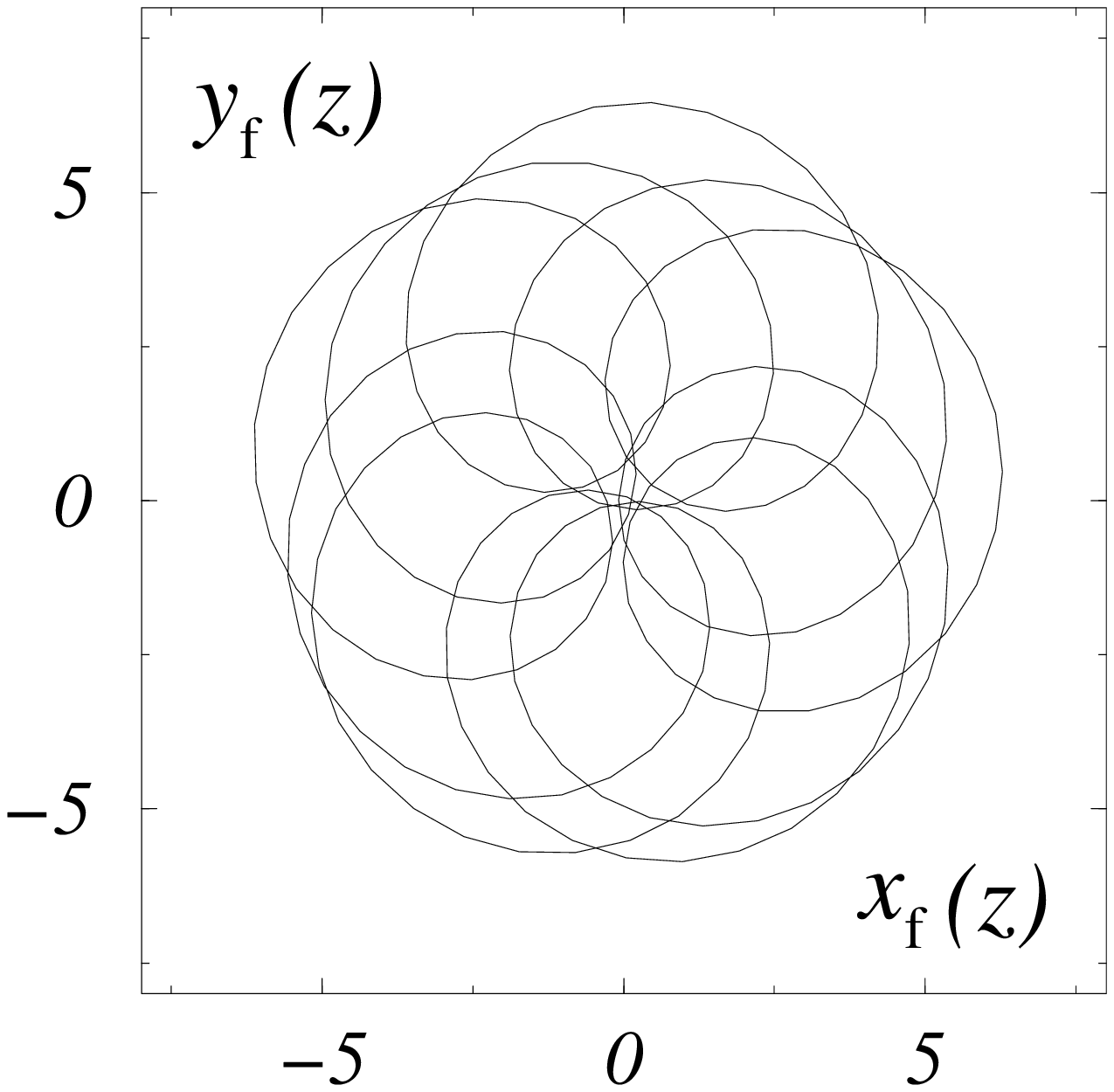}

\vspace{-0.6cm}

\centerline{\epsfxsize=5.3truecm
\epsffile{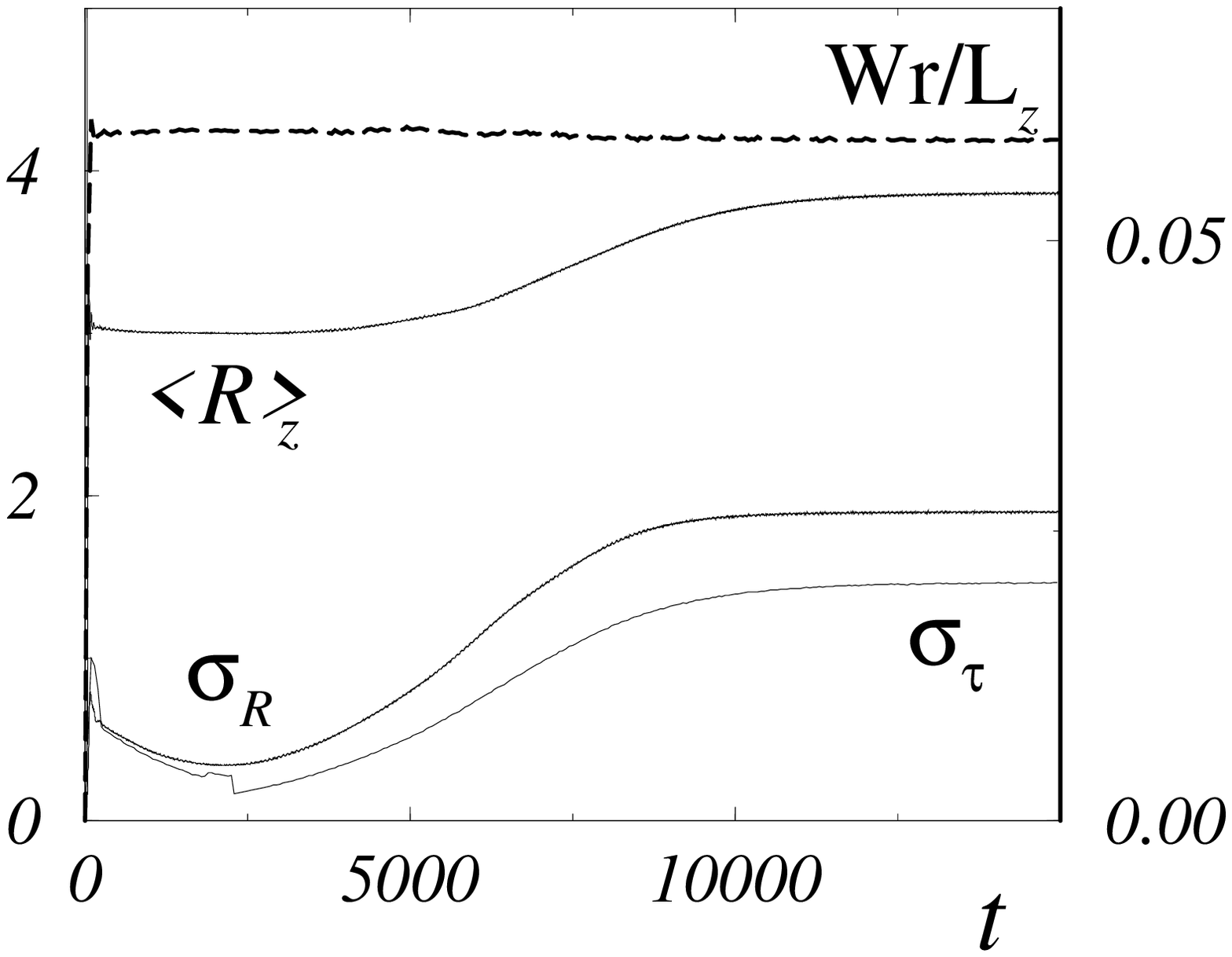}
\hspace{-.8cm}
\epsfxsize=5.3truecm
\epsffile{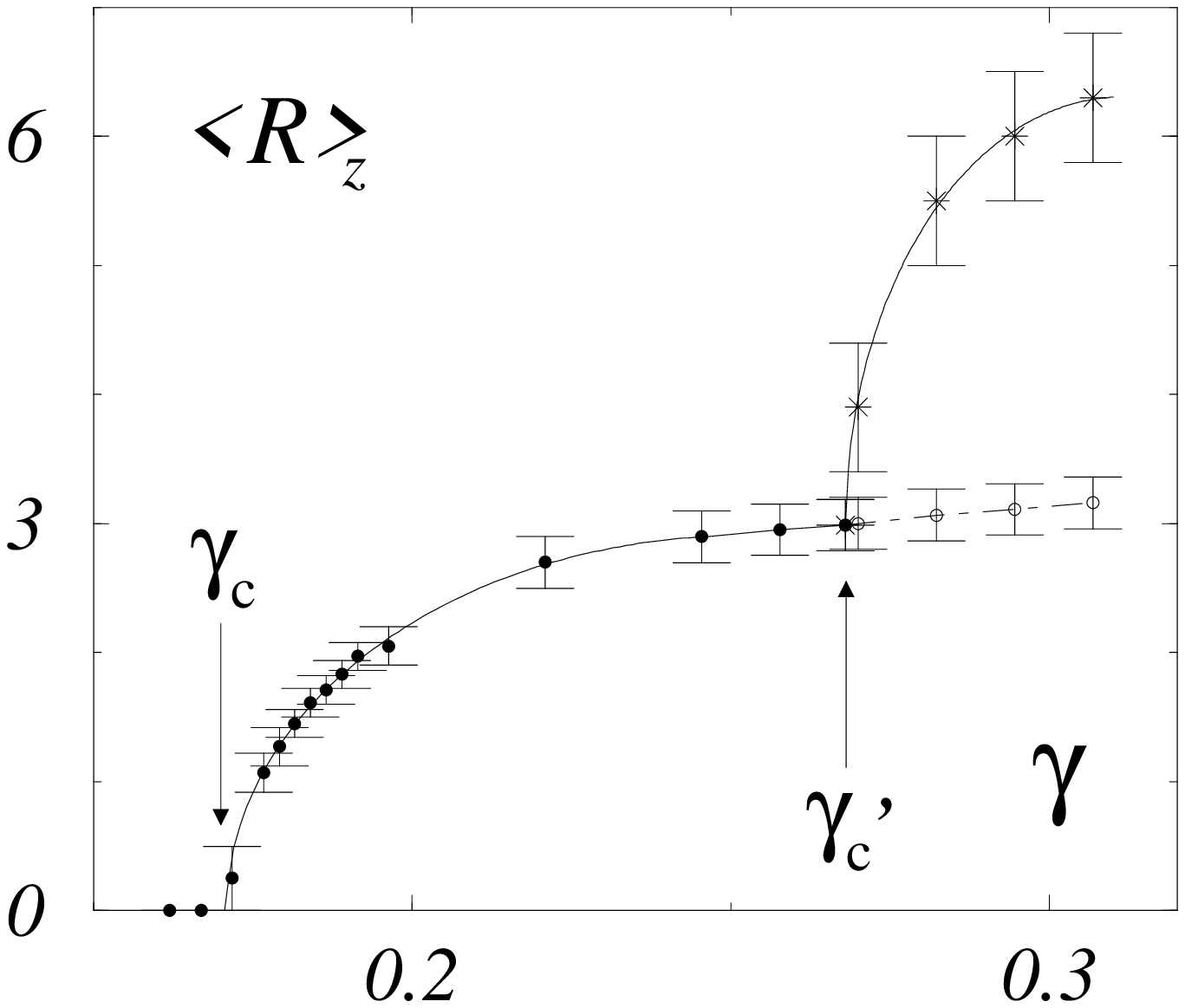}}
\vspace{-0.2cm}
\caption{Secondary Hopf bifurcation for $\alpha=1$ and $\beta=-0.5$.
This data was obtained under the contraint that ${\rm Tw}_{\rm r}=0$,
a small perturbation here, since $\gamma_{\rm h} \simeq \gamma$ 
for these parameter values.  
Top: $\gamma=0.27$; 3D isosurface $|A|=0.6$ colored by the 
phase field ($2\pi$-periodic grey scale);
Middle left: space-time plot of $\phi(z,t)$ (see Fig.~\ref{f5}).
Portion of size $L=600$ of a box of
length $L_z=2048$ shown during $\Delta t = 1825$ 
with an initial straight filament 
($L_x=L_y=92$, $\gamma=0.295$).
$2\pi$-periodic grey scale ($0=$black, $\pi=$white), 
except for the points where 
$\partial_z\phi< 0$ which are plotted in black regardless of the value
of $\phi$. This enhances the ``petals'',
or small loops, in the superhelix,
since only points of these 
loops have negative $\partial_z\phi$. For the primary helix, observed
as a transient, $\partial_z \phi = \gamma_{\rm h}>0$ everywhere. 
Wavelengths $\lambda_{\rm h}$, $\lambda_1$,  and $\lambda_2$ 
and velocities $v$, $v_1$ and $v_2$ are indicated. 
Middle right: projection of superhelix on the $(x,y)$ plane. 
The flower pattern is periodic due to the finite box length. 
Bottom left: $\gamma=0.27$: time series of $\langle\! R \rangle_z$ 
and $\sigma_R$ its rms over $z$ (left scale),
${\rm Wr}/L_z$ and $\sigma_{\tau}$ the rms of the torsion (right scale).
Bottom right: $\langle\! R \rangle_z(\gamma)$ 
across the two bifurcation points $\gamma_{\rm c}\simeq 0.171$ and 
$\gamma'_{\rm c} \simeq 0.267$. Dashed line: $R_{\rm h}$ for the (weakly)
unstable helices.
}
\label{f4}
\end{figure}

Moving away from the $\gamma_{\rm c}(\alpha,\beta)$ surface, 
the filament undergoes a secondary instability at some critical value
$\gamma'_{\rm c}> \gamma_{\rm c}$: a long-wavelength
modulation appears on the stable regular helix, eventually
producing a supercoiled
filament (Fig.~\ref{f4}). No reconnection event is observed. 
This structure is characterized by two 
well-defined wavelengths, $\lambda_1$ and $\lambda_2$, with
$\lambda_1 \sim \lambda_{\rm h}$ and  $\lambda_2 \gg \lambda_{\rm h}$ 
(Fig.~\ref{f4}, middle left). In general, for an infinite filament, 
the two wavelengths are incommensurate 
and the superhelix is a quasiperiodic object whose projection on the
$(x,y)$ plane produces a characteristic ``flower'' pattern (middle
right). Associated with these wavelengths are two frequencies, or 
two velocities $v_1$ and $v_2$ (middle left). 

The secondary instability is also a supercritical Hopf bifurcation.
Its forward nature and the nonlinear saturation are
apparent in the time evolution of an initially straight filament, which
quickly turns into a weakly unstable helix before bifurcating toward
a superhelix (Fig.~\ref{f4}, bottom left). 
Varying $\gamma$, the mean radius $\langle\! R \rangle_z$ is continuous 
but not differentiable at $\gamma'_{\rm c}$, and
the amplitude of the secondary modulations can be 
measured by $\langle\! R \rangle_z - R_{\rm h}$, 
where $R_{\rm h}$ is the radius
of the (unstable) primary helix (measured, e.g., from the transient helix 
stage) (Fig.~\ref{f4} bottom right).
As expected for a forward Hopf bifurcation, near threshold our data
is consistent with 
$(\langle\! R \rangle_z - R_{\rm h})^2 \sim (\gamma-\gamma'_{\rm c})$.
Even though the amplitude of this secondary instability saturates, 
we cannot rule out a chaotic asymptotic behavior.

Our measurements (from Fourier spectra) 
show that $1/\lambda_{\rm h} =1/\lambda_1 + 1/\lambda_2$,
at least for $\gamma-\gamma'_{\rm c}$ not too large.
Since the link number Lk is constant
(no reconnection occurred), this relation implies
that the ribbon component of the twist ${\rm Tw}_{\rm r}$ varies smoothly
across the transition.  \cite{TABOR}
This is corroborated by direct measurements of ${\rm Tw}_{\rm r}$ and 
the other topological quantities (Wr, ${\rm Tw}_{\rm f}$) 
across the secondary instability. 
During the evolution shown in Fig.~\ref{f4} (bottom left),
${\rm Wr}(t)$ is constant
through the second morphological change (bottom left) since 
 the mean torsion remains constant; $\tau(s)$, however,
becomes $\lambda_1$-periodic. Finally, the available data
does not allow to decide
whether these quantities still vary linearly with $\gamma$.

Although the instabilities studied here are rather 
traditional Hopf bifurcations, they are topologically constrained.
The existence of the ribbon component of the twist, 
particular to the filaments studied here, comes into play
in a somewhat ``passive'' way. Associated to the soft phase mode, it can
be seen as accomodating the torsion and writhe of the bifurcated structure.
Indeed, although it is a dynamical object 
supporting propagating waves, our filament coils and supercoils 
under twist like 
an elastic rod or a DNA molecule 
(see \cite{TABOR,ROD} and references therein). 
Moreover, the simple behavior of Wr 
supports the adoption  a topological viewpoint 
when studying  the dynamics of 3D phase singularities. 
We note, however, that the ribbon twist might play a more important role
in excitable media or systems with complex oscillations, for which
the gauge invariance of the phase is lost.

\end{multicols}


\begin{thebibliography}{99}
\bibitem{NOTE1} With no-flux conditions in $z$, the link
would not be a conserved quantity, complicating further the picture.
\end{thebibliography}
\end{document}